# Experimental investigations of acoustic curtains for hospital environment noise mitigations

Sanjay Kumar, Rui Qin Ng, and Heow Pueh Lee

Department of Mechanical Engineering, National University of Singapore, 9 Engineering Drive 1, Singapore 117575, Singapore

Corresponding author's email: mpesanj@nus.edu.sg (S.K.); mpeleehp@nus.edu.sg (H.P.Lee)

**Abstract**

The continuous increase of hospital noise levels has become a vital challenge for society. The complex soundscapes in the hospital produce unpleasant noise, which may exceed the prescribed noise level for the patients and healthcare professionals. Previous studies have reported that extended exposure to loud noise may cause auditory and nonauditory disorders in healthcare professionals, medical staff, and patients. Therefore, there is an increased interest for the design and fabrication of effective noise barriers for the hospital premises. Herein, we have performed the thorough experimental investigations on the acoustical performances for PVC coated polyester fabrics and 100 % pure PVC sheets. The performances of these potential acoustic curtains have found to be superior to that of existing acoustic curtains for hospitals. Also, the results showed that the sound transmission class rating of PVC curtains are much higher than the existing commercial acoustic curtains.

Keywords: Hospital noise; acoustic curtains; noise mitigation; sound transmission loss, sound transmission class.

## 1. Introduction

In the past few decades, the hospital noise has increased substantially and posed a considerable risk to the hospital occupants and commuters. There are several notable noise sources in hospital premises such as in intensive care units (ICUs) - the array of buzzers, pagers, telephones, alarms, beeps, the conversation among medical personals, patients, and family members; and in general wards: movement of surgical trolleys besides verbal conversation. Together, this generated noise can create a highly unpleasant and unhealthy environment for doctors and nurses, and also for critical care patients. The background noise level in the hospitals generally ranges from 50 dB(A) to 70 dB(A) during daytime hours, and during the peak hours, it often exceeds 85 dB(A) to 90 dB(A). Some medical equipment, like portable X-ray machines, produces noise levels much higher than 90 dB(A) [1, 2]. The World Health Organization (WHO) has considered a high noise level as one of the most crucial factors having a direct adverse impact on the health of medical professionals and patients [3]. Previous research findings support WHO's claims that the prolonged exposure to these unpleasant noises with exceeding limits may induce negative physiological and psychological impacts on patients and health professionals. The prolong exposure to noise in the hospitals has severe implications for patients such as sleeping disruption [4, 5], hypertension, agitation, annoyance, gastrointestinal system problems, nervousness, psychic disorders, cardiac arrhythmia, elevated blood pressure, increased heart, and respiration rate among intensive care patients. Besides, among healthcare professionals and medical staff exposed to loud noise, several psychological complications have been reported, like increased perceived work pressure, prone to committing work errors, stress, annoyance, headache, increased fatigue levels, emotional exhaustion, and burnout. In recent years, the World Health Organization (WHO) has released specific guidelines to address the problem of hospital noise. The set values for continuous background noise in hospital patient rooms are 35 dB(A) during the day and 30 dB(A) at night, with maximum peaks (night-time) in hospital wards not to exceed 40 dB(A).

In addition to several adverse effects on patients and healthcare professionals, two other important factors, namely, speech intelligibility and speech privacy, are directly affected by high noise levels. Speech intelligibility is referred to as the clarity of speech communications among participants, while speech privacy is related to conversation confidentiality. For the avoidance of possible low-to-severe casualties, if any, the directives of the lead doctors should be conveyed among other professionals during the operation or examination of patients. Similarly, the conversation between patients and healthcare professionals should be private and confidential. In 1996, the Health Information Portability and Accountability Act (HIPAA) came into the effect to provide patient confidentiality in US. This act has further elevated to provide reasonable safeguards to the patient's health information provided either in written form or orally (Department of Health and Human Services, Office for Civil Rights, US, 2003).

Considerable efforts have been made by the researchers for the assessment of hospital noise and their control. However, most of the reported studies are observational and descriptive [6-13]. They have mainly focused on the identification of hospital noise problems and assessment of their impact on human health consequences. These studies were mostly conducted by medical professionals and reported in the medical literature. The control of hospital noise had been bound to administrative control measures, with few engineering strategies. Therefore, a holistic approach is much needed towards the design and development of an effective system for noise reduction in crucial work places like hospitals. In recent years numerous interventions have been attempted for the reduction of hospital noise for providing a healthy work environment. Key architectural design solutions include installations of high-performance sound-absorbing panels [14], upgradation of medical equipment, i.e., noiseless paging systems [1], use of earmuffs/earplugs [15-18], and installation of the acoustic curtains to create a partition between adjacent patient beds [19]. Among these, implementation of the acoustic curtains is most beneficial because of several reasons such as quick installments, reconfigurable, washable, and lower cost. Some widely used materials for the fabrication of curtains are woven fabrics [20, 21], cottons [22], natural fibers [23], and non-woven polymers [24]. These acoustic fabrics have shown excellent capability for noise mitigation applications. However, the hospital curtains are quite different from the ordinary acoustic curtains. The curtains used in the hospitals could be contaminated with pathogenic bacteria. The acoustic curtains should possess some specific properties like fire-retardant, washable, light-in-weight, and contamination-proof from healthcare-associated pathogens. Recently, Ohl et al. [25] performed a longitudinal study to assess the time period and persistence of bacterial contamination on the several hospital privacy curtains. Al-Tawfiq et al. [26] experimentally investigated the antibacterial characteristics of privacy curtain. The curtain was made of non-woven polypropylene materials and treated with Fantex. Shek et al. [27] performed an observational study to determine the contamination rate of hospital privacy curtains. It was revealed that most of the curtains were contaminated with bacteria after fourteen days of their use.

Herein, we performed the experimental investigations on the acoustical performances of various acoustic curtains made of Polyvinyl chloride (PVC), polystyrene, and Polypropylene. Several strategies supported by the concepts of acoustical physics have been utilized for the selection of the curtains. The acoustical performance of these curtains has been comprehensively investigated on small-scale and large-scale levels by standard impedance tube and reverberation room methods, respectively. A comparison has also made between the PVC curtains and the commercially available curtains. The sound transmission class of all the curtains has determined and presented for their significance.

## 2. Materials and methods

### 2.1 Materials

The materials for the fabrication of hospital noise barriers were carefully selected. Five different materials were procured and investigated; namely, PVC coated polyester (PE) fabric (Feicheng Hengfeng Plastic. Co), Standard Hospital Curtain (Shaoxing Dairui®), pure PVC sheet, TANGO Acoustical Curtain, and Polypropylene sheet (Nippon Home).

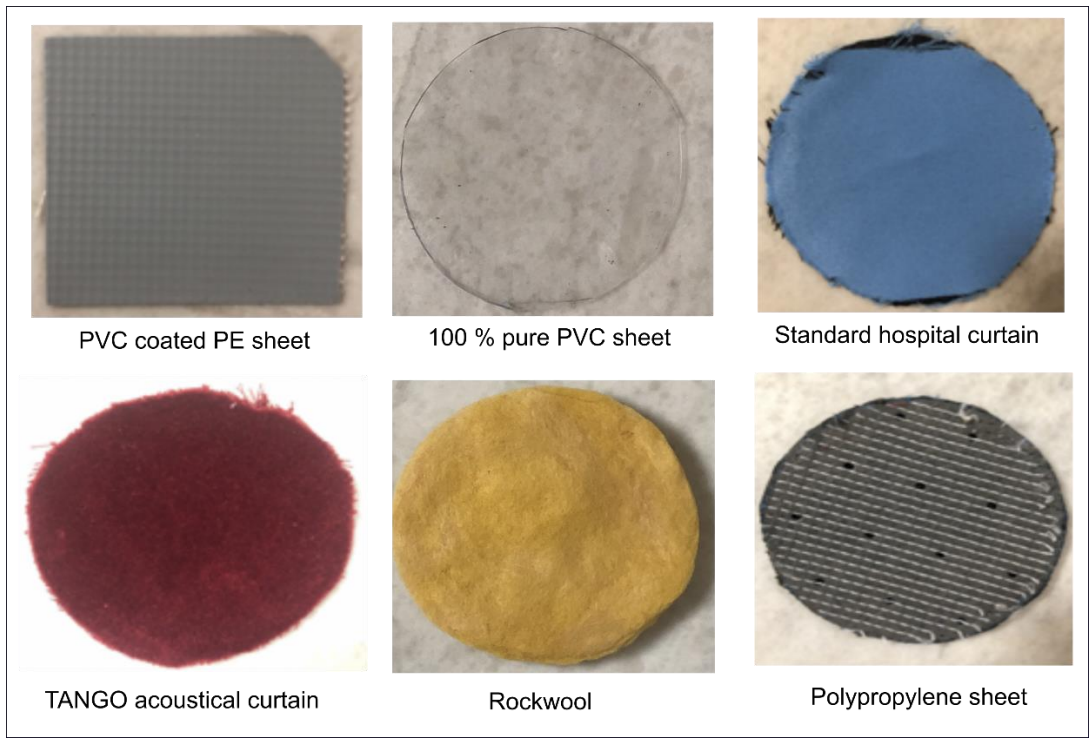

**Figure 1.** Photographs of the materials selected for the experimental investigations.

### 2.2 Acoustical measurements

#### 2.2.1 Impedance tube method

A standard sound impedance tube (BSWA®) was used for the measurements of transmission loss. The specifications of the impedance tube were designed as per the guidelines given in the British standard EN

ISO 10534-2:2001 [28]. **Figure 2**(a) shows the schematic diagram of the four-microphone impedance tube system. For the transmission loss measurements, a four-microphone setup was adopted where two pairs of microphones (BSWA MPA416) were installed at the upstream and downstream of the test specimen to acquire sound waves pressure data. For mounting of thin and flexible curtains in the impedance tube, a 3D printed polylactide ring-shape sample holder (outer dia. 99.8 mm, inner dia. 79 mm, and thickness 14 mm) was used. The test specimens of diameter 99.8 mm were attached on the sample holder by using the double-sided adhesive tape. To ensure uniform tension of the specimens, these were laid flat before affixing to the contacting surface of the holder, as suggested by Weilnau et al. [29]. Moreover, the orientation of the installed test sample holder inside the tube was changed across various measurements to minimize the possible experimental errors by sample placement.

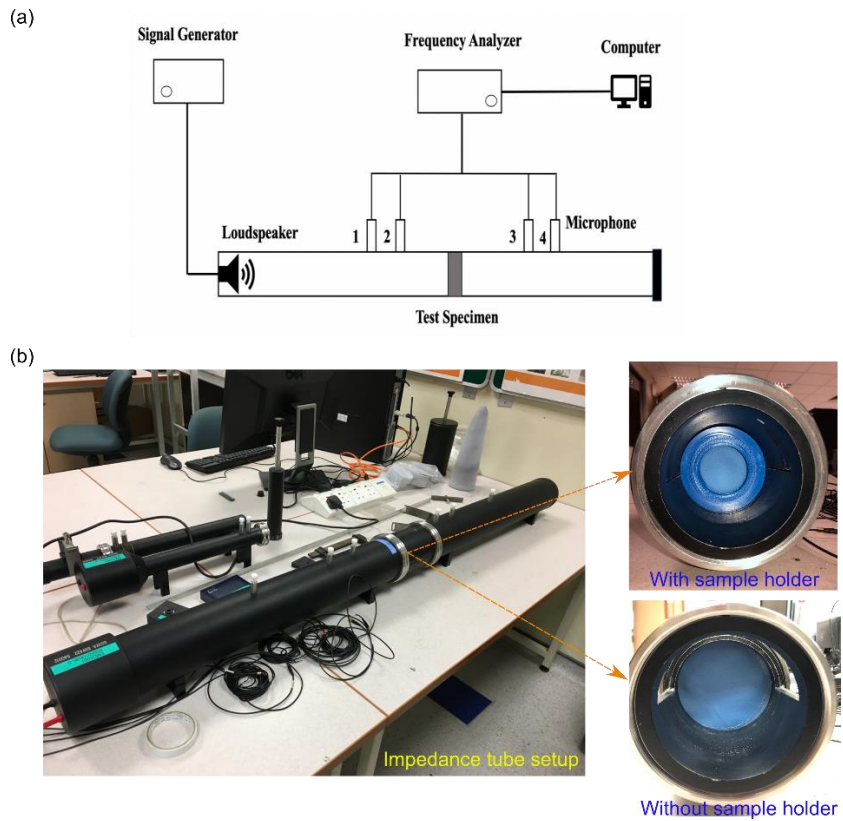

**Figure 2.** Sound frequency spectrogram of recorded hospital wards audio.

### 2.2.2 Reverberation room method

The acoustical performance of large-scale specimens was performed in the ASTM standard reverberation chamber at the National University of Singapore acoustics laboratory. The chamber consists of a source and receiver room having volumes of 65.5 m$^3$ and 81.1 m$^3$, respectively. The difference in room volumes is about 19.2%, fulfilling the volume difference requirement of at least 10%, as specified in ASTM E90 [30]. The surface areas of the source and the receiver chambers were 97.2 m$^2$ and 109.5 m$^2$, respectively. The rooms were separated by a partition wall with a central opening (0.995 X 0.995 X 0.104 m$^3$) for mounting of specimens. The temperature and relatively humidity on the day of testing was 23.7°C and 66.3%, respectively. Test specimens were tight fitted carefully within the central opening area to ensure there would be no sound leakages. Two wooden frames were used to firmly hold the thin curtains ensuring the flatness of these specimens. As the specimens had a smaller thickness relative to the thickness of the partition wall opening, wooden frames were designed as a filler for the specimens to be held firmly by the pre-installed toggle clamps. **Figure 3**(a) shows the photographs of the PVC coated curtain installation setup for sound acoustical measurements. Gaskets were placed in between the contacting surfaces of the wooden frames and toggle clamps to minimize potential sound leakages.

*Reverberation time*

The room reverberation time, T60, is defined as the time taken for the sound level to decay 60 dB after the sound source has stopped. The interrupted noise method was adopted for calculation of T60, whereby pink noise was generated from an external noise source (Brüel & Kjær Type 2734) and amplified via an omnidirectional amplifier (Larson Davis BAS001). The minimum difference between the generated noise and background noise levels was ensured at least 20 dB. The noise generator was then switched off once the noise level had stabilized, and the corresponding decay time for 20 dB (T20) was measured using a hand-held sound level meter (Larson Davis 831). The decay time T20 was measured at three different positions with each having a total of five measurements, and average values were taken for the final calculation. The measuring positions (**Figure 3**b) were selected as per the ASTM 2235 standard [31]. Also, the sound level meter was held at least 0.5 m away from the body to avoid any sound reflections from it. It was difficult to stimulate a diffuse noise field with a magnitude large enough so that the noise could decay by 60 dB with the available sound generator. So, the ensembled reverberation time, T60, was determined under the assumption of linear decay rate of 20 dB multiplied by 3. Whilst, the decay rate $d$ was determined by $d = 60/T60$. Consequently, the room sound absorption was computed by $A_r(f) = 0.921 \times \frac{V\,d}{c}$, where $A_r(f)$ denotes absorption of the receiving room in m$^2$, V denotes the volume of the receiver room, and c denotes the speed of sound $c = 20.047\sqrt{t_{room}(Kelvin)}$.

*Sound transmission loss calculation*

A separate set-up was arranged for the measurement of average sound pressure levels (SPLs) in the source room $L_s(f)$, and in the receiving room, $L_r(f)$ as per ASTM E90. **Figure 3**(c) shows the schematic representation of the reverberation room setup for the SPL measurement. A continuous pink noise was played into the source room using the two omnidirectional loudspeakers (Yamaha DXR15), driven by separate random noise generators and amplifiers. The loudspeakers were placed at the trihedral corners of the room to excite room modes effectively. Two calibrated microphones in the source room and the receiver room were used to capture the sound levels in the respective rooms. All the measurements were performed in all one-third-octave bands with nominal mid-band frequencies specified in ANSI S1.11 [32] from 100 to 5000 Hz. A total of six measurements were performed for each specimen, and their average values were considered in transmission loss calculation. The sound transmission loss (STL) was evaluated by the following relations $STL = L_s(f) - L_r(f) + 10\,\log_{10}(S/A_r(f))$, where $S$ denotes the surface area of the test specimen. Besides, repeatability test was performed at each frequency to check with required confidence interval for transmission loss measurements as suggested by standard ASTM E90 [30].

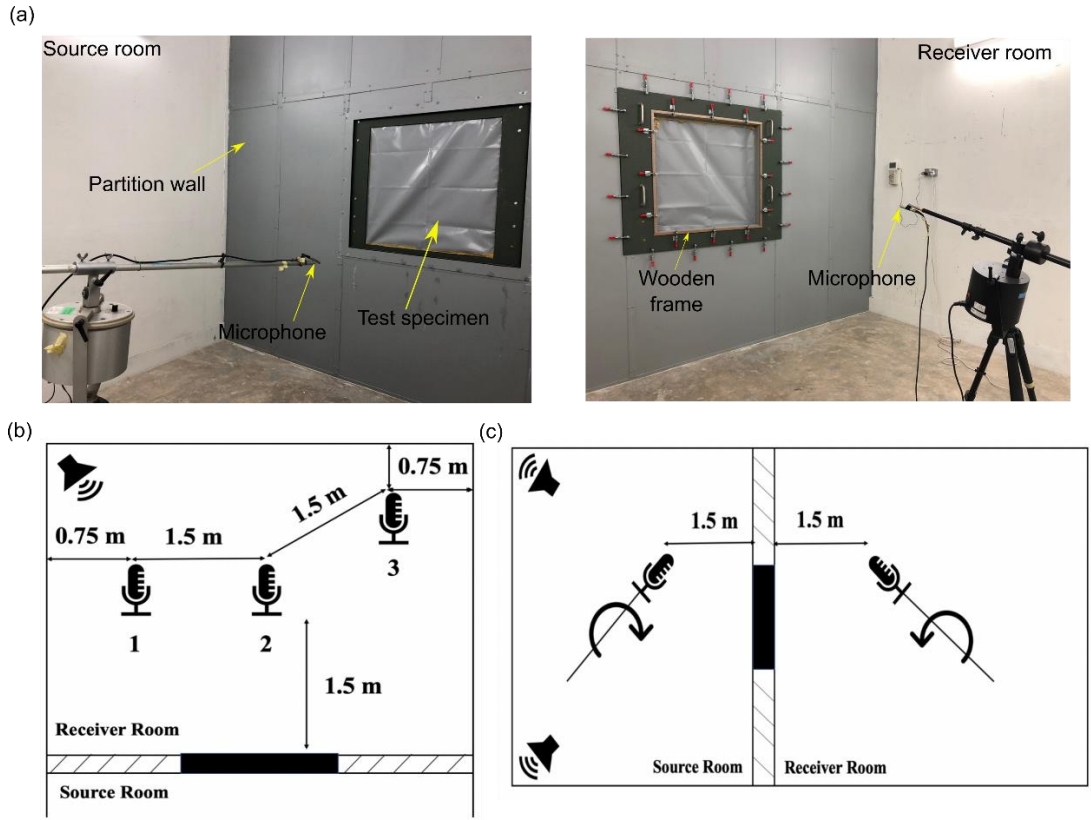

**Figure 3.** (a) The photographs of the installed test specimen for acoustical measurements. Schematic representation of experimental setup for the measurements of (b) decay time T20 and (c) sound pressure levels.

## 3. Results

**3.1 Hospital sound level measurement**

**Figure 4** shows the typical average sound levels at two different locations general ward and trauma centers in Hospitals. For both settings, conversations and staff activities were distinguishably noted as the primary source of generated noise within wards. In terms of SPL, both curves displayed relatively good agreement in the overall trend, with higher noise levels observed between 315 Hz and 6300 Hz and a peak level at about 1000 Hz. The average sound level in the general ward was higher than 45 dBA in 250 Hz-5500 Hz, which far exceeds the recommended suggested values (i.e., 45 dBA during daytime) by the United States Environmental Protection Agency (USEPA) [33]. Loupa et al. [34] reported a similar noise spectrum with the SPL peaks between 500 Hz and 2000 Hz for the hospital pulmonary departmental wards. Based on these measured noise level trends, we target the noise reduction in low-to-mid frequency (250-1250 Hz).

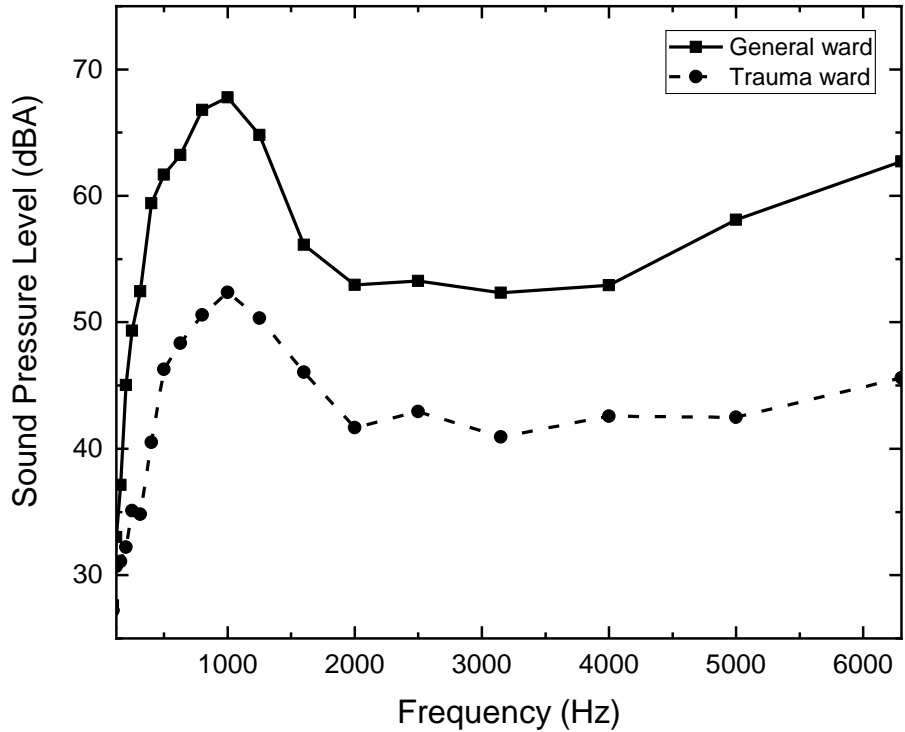

**Figure 4.** Sound frequency spectrogram of recorded hospital wards audio. Similar peak frequency range have been reported by Livera et al. [35] and Loupa et al. [34].

## 3.2 Sound transmission loss measurement (small-scale)

Before the sound transmission loss measurement of the acoustic curtains, the effect of the sample holder was evaluated first. The sample holder is required for the ease of mounting the test samples. In the process, the transmission loss of the acoustic curtain was measured under two different conditions; with specimen holder and without specimen holder. In the presence of the sample holder, around 1 to 2 dB increments in the transmission loss values were observed over 125-1600 Hz (see **Figure 5**). As shown, the sample holder had a negligible effect on the measured STL values for the test specimens. Therefore, all the impedance tube measurements were performed using the sample holder. In the procedure, the position of sample holder was varied by rotating it along the circumference of the tube in an interval of 45 degrees and STL was measured at each position. Further, to minimize the effect of sample mounting errors, five repetitions of STL measurements were taken at each position, and the average of these values was used in the study. **Figure 6**(a) shows the transmission loss spectra of the various test specimens. The standard hospital polyester curtain displayed relatively poor STL performance of about 5 dB in 400-1600 Hz range while the PVC coated PE fabric displayed a better performance, particularly in the mid-to-high frequency range with a transmission loss of about 15 to 22 dB. The significant difference in performance between these two materials is attributed to the material surface properties. A thin uniform coating of PVC on the PE sheet provides a smooth, impermeable surface from which incoming sound waves may be reflected, thus resulting in reduced sound transmission. Moreover, the viscoelastic nature of PVC coating provides enhanced damping of the sound vibrations, whereas, for the standard fabric curtains, the sound wave is partially transmitted through the fabric given its porous nature. Besides, the higher relative surface density of PVC coated PE curtain may be another reason for its superior transmission loss performance.

To get better insights of acoustical characteristics of PVC-based curtains, the transmission loss of 100 % pure PVC sheet was measured. **Figure 6**(b) depicts the measured STL values for these curtains. The results revealed that the combined effects of PVC and polyester resulted in a better performance than pure PVC curtains.

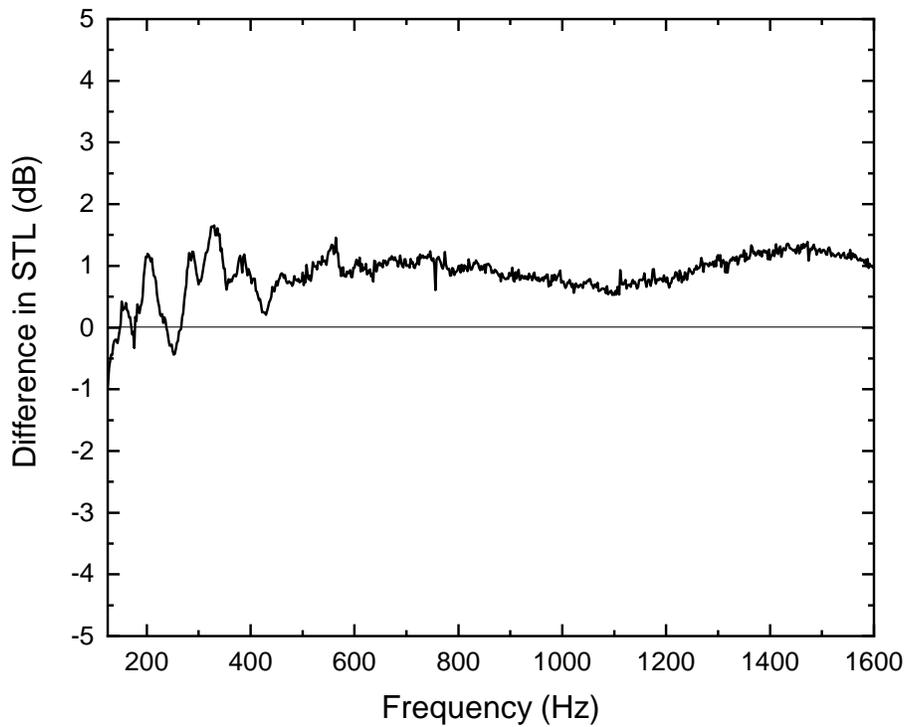

**Figure 5.** Influence of sample holder on transmission loss performance of acoustic curtain.

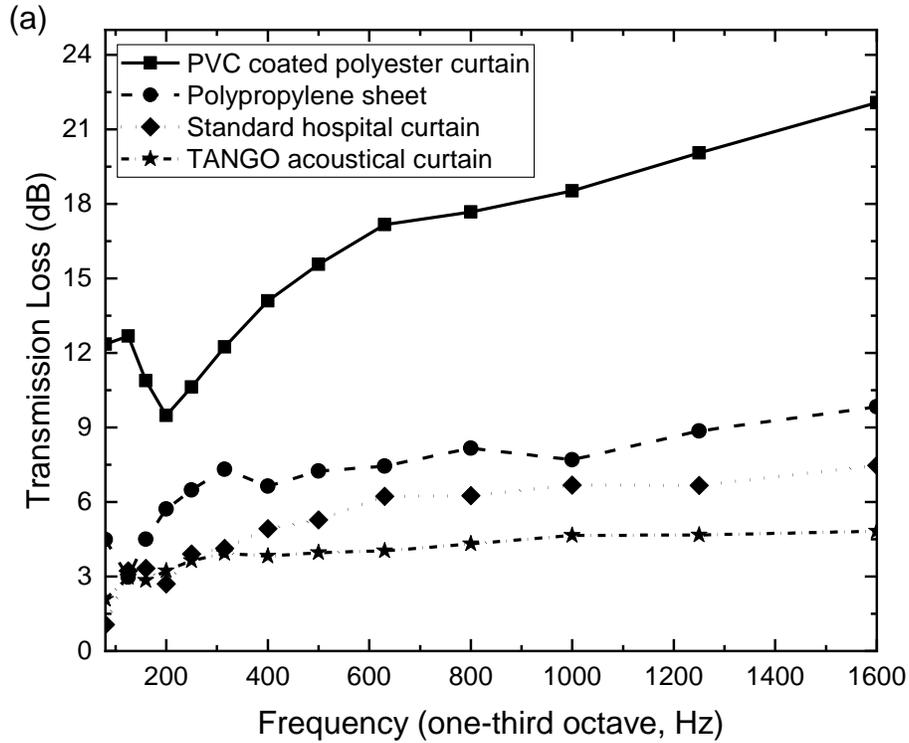

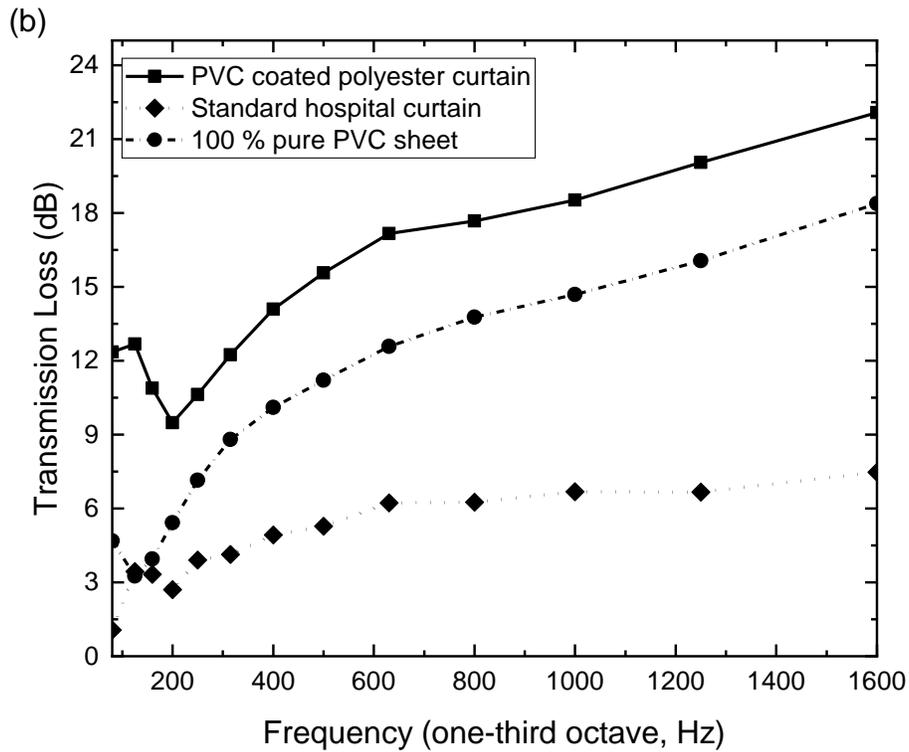

**Figure 6.** Transmission loss performance of various test specimens based on impedance tube test. (a) PVC coated polyester curtain, propylene (PE) sheet, standard hospital curtain, and TANGO acoustical curtain. (b) PVC coated polyester curtain, standard hospital curtain, and 100 % pure PVC. The results were plotted after the exclusion of STL values for the sample holder.

### 3.3 Sound transmission loss measurements (large-scale)

Since the acoustical performance of PVC coated polyester and pure PVC curtains are far better than the other materials (see **Figure 6**). These two materials were further investigated using the reverberation room method. The performance of both the curtains was evaluated in two different noise source settings viz. standard pink noise and emulated hospital ward noise. **Figure 7**(a) and **Figure 7**(b) shows the measured sound transmission loss spectra for both noise source settings. As shown, the STL values obtained in the presence of pink noise are in good agreement with values obtained using pink noise.

Further, the STL performance of the PVC coated PE fabric, and pure PVC sheet curtain showed a good agreement with each other. Also, the STL value increased with the increment of frequency values, following the mass-frequency law. The trend was consistent with the results reported by Wang et al. [36] for the STL performance of PVC composite materials. Further, an STL peak value of around 15 dB can be observed in the low-frequency zone (< 500 Hz). Such a high transmission loss peak value achieved from the PVC-based curtains are auspicious and may find a potential application for low-to-mid frequency noise insulation.

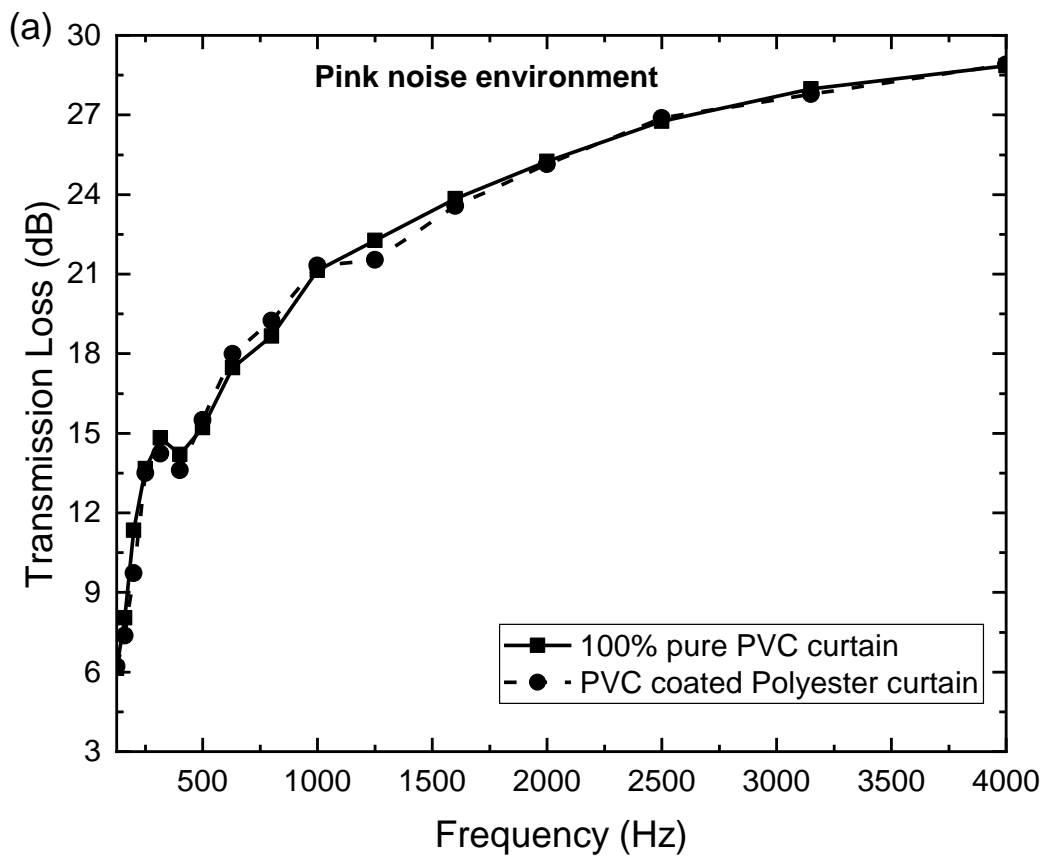

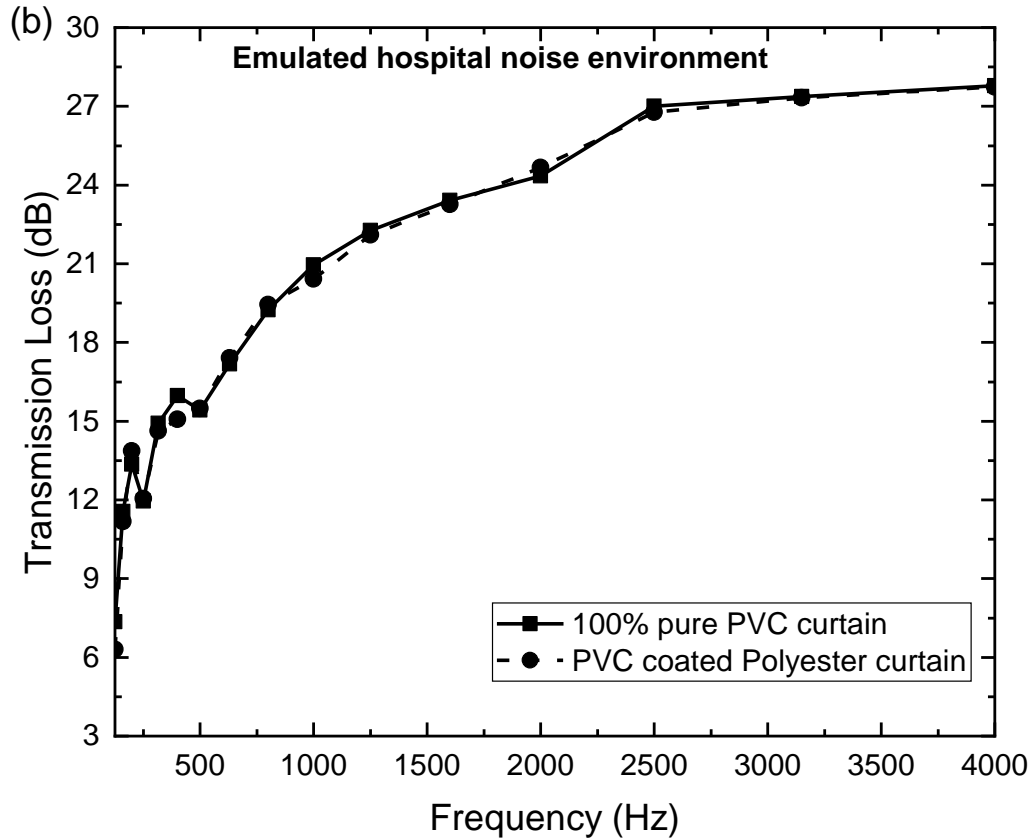

**Figure 7.** Sound transmission loss performance of 100 % pure PVC and PVC coated polyester curtains in two different noise environments (a) Pink noise, and (b) emulated hospital noise.

### 4. Discussion

In this study, the transmission loss of the curtains was measured using two experimental methods, the impedance tube method, and the reverberation room method. As presented earlier section, the PVC coated PE curtain and 100 % PVC curtain exhibit outstanding sound barrier performances. For impedance measurement, the average STL values of PVC coated PE curtain is higher than that of 100 % PVC curtain. However, the reverberation room calculated STL values are almost similar for both curtains in pink noise and emulated hospital noise settings. The discrepancy in the STL spectrum for the impedance tube and the reverberation chamber methods are attributed to the sound source field. In the impedance tube method, an incident one-dimensional plane wave is used as an input source, and the anechoic end is used to minimize the backscattered wave in the downstream side of the tube. Whereas, for the reverberation room methods, a diffuse sound field is generated in the source room, and the reverberation of the receiving room is included in the calculation. The diffused sound field is a more realistic emulation of actual ambient condition. Therefore, the reverberation room method provides practical and feasible information about the acoustical performance of test specimens.

For a better assessment, the acoustical performance of the proposed PVC curtains was compared with the existing commercial products. The reverberation chamber method was used for sound transmission loss measurements for all three commercial curtains, and the respective STL values were subtracted from STL values obtained for PVC coated PE curtain and 100 % pure PVC curtain. **Figure 8** (a) and (b) show the net sound transmission loss spectra of acoustic curtains compared with PVC coated PE curtain and 100 % pure PVC curtain, respectively. The positive values in the plots denote the PVC curtains fared better than the commercial products and vice versa.

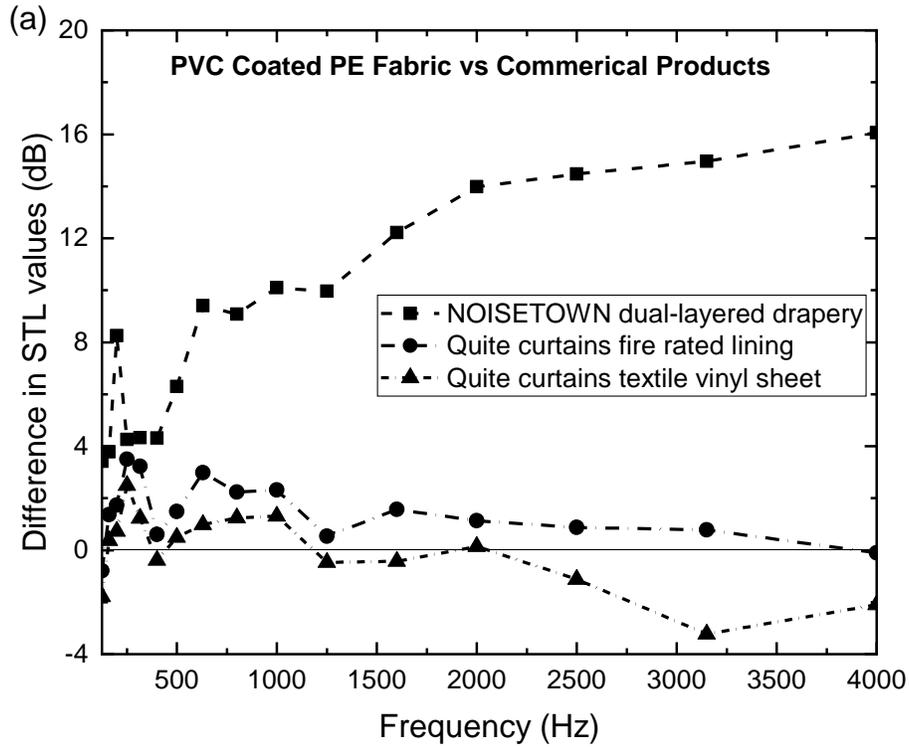

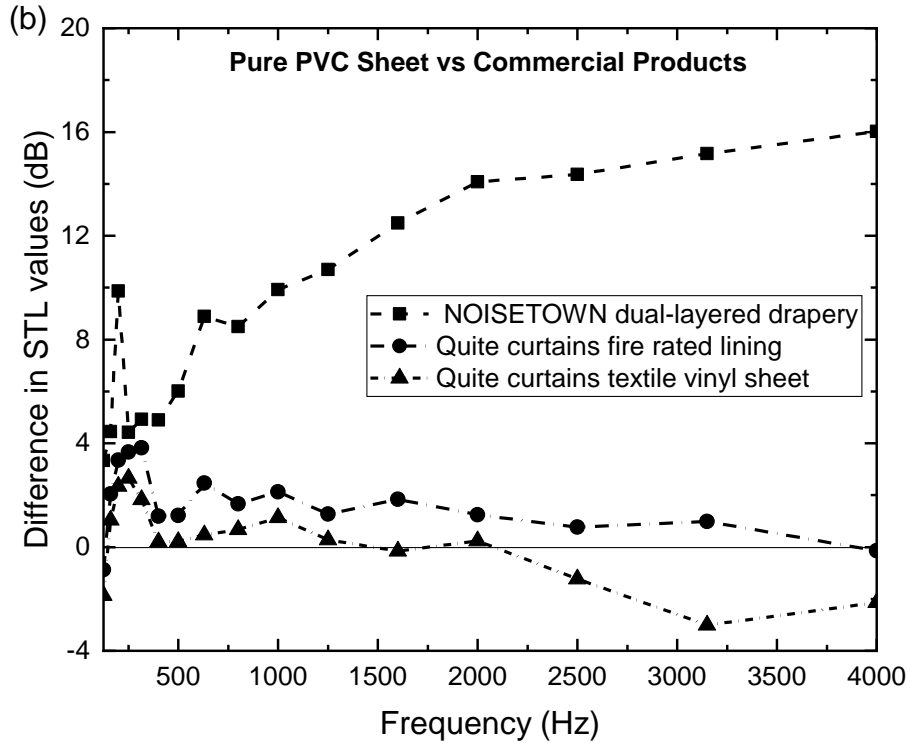

**Figure 8.** Sound transmission loss performance comparison of (a)100 % pure PVC and (b) PVC coated polyester curtains with commercial curtains.

For the more realistic acoustical evaluations, the sound transmission class (STC) value was computed for all curtains. STC is a single-number standard rating of materials or composite's sound insulation capabilities. The STC is a weighted average of the sound transmission loss values at one third-octave band frequencies that are normalized using the area of the standard wall partition and the sound absorption in the receiving room [37, 38]. It is widely used to rate the sound isolation of interior partitions, ceilings/floors, doors, and windows. **Table 1** shows that the STC ratings of the proposed PVC curtains are higher than the commercial acoustic curtains. We achieved an STC of 21 with the PVC-based curtains which is considered to be efficient for low-to-mid frequency noise mitigation.

In this study, we did not perform the anti-microbial testing of the proposed PVC based-curtains. Previous studies have revealed the anti-bacterial characteristics of Polyvinyl chloride-based curtains [25, 39]. Soft, smooth, and hydrophobic qualities of PVC curtains were found to result in lower infection rates and reduced risk of pathogen transmission compared with standard privacy curtains.

**Table 1.** STC rating of PVC curtains and other commercially available curtains.

| Product description | STC Rating |
|---|---|
| NOISETOWN dual-layered drapery | 13 |
| Quite curtains fire rated lining | 17 |
| Quite curtains great lakes textile vinyl sheet | 20 |
| **PVC coated polyester fabric** | **21** |
| **100 % pure PVC sheet** | **21** |

## 5. Conclusion

In summary, we experimentally investigated the noise mitigation characteristics of PVC-based curtains. The transmission loss of these curtains was measured using two experimental methods, the impedance tube method, and the reverberation room method. The experimental results have shown superior sound barrier performance of PVC curtains compared with that of the commercially available hospital curtains. The STC rating of PVC curtains is higher than that of other commercially available curtains. The results of these investigations imply that the existing low-performance privacy curtains can be replaced by the proposed collapsible, hydrophobic, washable, and opaque PVC coated PE curtains. The transparent PVC coated sheets have a great potential to be used as an acoustical window curtain to reduce outdoor road traffic noise while enabling daylight transmission into the rooms.

**Acknowledgement**

The first author would like to acknowledge the financial support from the Ministry of Education RSB Research Fellowship, Singapore.

**Author contributions**

Sanjay Kumar and Rui Qin Ng have equal contribution in the manuscript.

**Conflicts of Interest**

The authors declare no conflict of interest.

**References**

[1] A. Joseph, R. Ulrich, Sound control for improved outcomes in healthcare settings, The Center for Health Design, 4 (2007) 2007.
[2] H. Xie, J. Kang, G.H. Mills, Clinical review: The impact of noise on patients' sleep and the effectiveness of noise reduction strategies in intensive care units, Crit. Care, 13 (2009) 208.
[3] S. Kumar, H.P. Lee, The Present and Future Role of Acoustic Metamaterials for Architectural and Urban Noise Mitigations, Acoustics, 3 (2019) 590-607.
[4] J.F. Schnelle, J.G. Ouslander, S.F. Simmons, C.A. Alessi, M.D. Gravel, The nighttime environment, incontinence care, and sleep disruption in nursing homes, J. Am. Geriatr. Soc., 41 (1993) 910-914.
[5] N.S. Freedman, J. Gazendam, L. Levan, A.I. Pack, R.J. Schwab, Abnormal sleep/wake cycles and the effect of environmental noise on sleep disruption in the intensive care unit, Am. J. Respir. Crit. Care Med., 163 (2001) 451-457.
[6] A. Konkani, B. Oakley, B. Penprase, Reducing hospital ICU noise: a behavior-based approach, J. Healthc. Eng., 5 (2014) 229-246.
[7] B.B. Kamdar, J.L. Martin, D.M. Needham, Noise and light pollution in the hospital: a call for action, J. Hosp. Med., 12 (2017) 861.

[8] B. Vuksanovic, R. Arias, M. Machimbarrena, M. Al-Mosawi, Monitoring and Analysis of Noise Levels in Intensive Care Units Using SSA Method, in: INTER-NOISE and NOISE-CON Congress and Conference Proceedings, Institute of Noise Control Engineering, 2019, pp. 7507-7518.
[9] J.M. Bliefnick, E.E. Ryherd, R. Jackson, Evaluating hospital soundscapes to improve patient experience, J. Acoust. Soc. Am., 145 (2019) 1117-1128.
[10] G. Loupa, Influence of Noise on Patient Recovery, Curr. Pollut. Rep., (2020) 1-7.
[11] E.R. Huisman, E. Morales, J. van Hoof, H.S. Kort, Healing environment: A review of the impact of physical environmental factors on users, Build. Environ., 58 (2012) 70-80.
[12] L. Walker, C.A. Karl, The Hospital (Not So) Quiet Zone: Creating an Environment for Patient Satisfaction Through Noise Reduction Strategies, Health Environ. Res. Des. J., 12 (2019) 197-202.
[13] D. Juang, C. Lee, T. Yang, M. Chang, Noise pollution and its effects on medical care workers and patients in hospitals, Int. J. Environ. Sci. Tech., 7 (2010) 705-716.
[14] P.M. Farrehi, B.K. Nallamothu, M. Navvab, Reducing hospital noise with sound acoustic panels and diffusion: a controlled study, BMJ Qual. Saf., 25 (2016) 644-646.
[15] L.K. Zahr, Premature infant responses to noise reduction by earmuffs: effects on behavioral and physiologic measures, J. Perinatol., 15 (1995) 448-455.
[16] C.J. Wallace, J. Robins, L.S. Alvord, J.M. Walker, The effect of earplugs on sleep measures during exposure to simulated intensive care unit noise, Am. J. Respir. Crit. Care, 8 (1999) 210.
[17] A. Richardson, M. Allsop, E. Coghill, C. Turnock, Earplugs and eye masks: do they improve critical care patients' sleep?, Nurs. Crit. Care, 12 (2007) 278-286.
[18] S. Sweity, A. Finlay, C. Lees, A. Monk, T. Sherpa, D. Wade, SleepSure: a pilot randomized-controlled trial to assess the effects of eye masks and earplugs on the quality of sleep for patients in hospital, Clin. Rehabil., 33 (2019) 253-261.
[19] D.S. Pope, E.T. Miller-Klein, Acoustic assessment of speech privacy curtains in two nursing units, Noise & health, 18 (2016) 26.
[20] X. Tang, D. Kong, X. Yan, Multiple regression analysis of a woven fabric sound absorber, Text. Res. J., 89 (2019) 855-866.
[21] R. Pieren, B. Schäffer, S. Schoenwald, K. Eggenschwiler, Sound absorption of textile curtains–theoretical models and validations by experiments and simulations, Text. Res. J., 88 (2018) 36-48.
[22] R. Atiénzar, M. Bonet, J. Payà, R. del Rey, R. Picó, Sound absorption of doped cotton textile fabrics with microcapsules, Revista de Acústica, 50 (2019) 13-21.
[23] C. Chen, M. Bi, J. Tang, B. Zhao, Z. Wang, G. Wang, Microstructure and acoustic property of windmill leaf sheath, fruit bunch and leaf fiber materials, Mater. Res. Exp., 6 (2019) 095108.
[24] P. Segura-Alcaraz, J. Segura-Alcaraz, I. Montava, M. Bonet-Aracil, The effect of the combination of multiple woven fabric and nonwoven on acoustic absorption, J. Ind. Text., (2019) 1528083719858771.
[25] M. Ohl, M. Schweizer, M. Graham, K. Heilmann, L. Boyken, D. Diekema, Hospital privacy curtains are frequently and rapidly contaminated with potentially pathogenic bacteria, Am. J. Infect. Control, 40 (2012) 904-906.
[26] J.A. Al-Tawfiq, A.M. Bazzi, A.A. Rabaan, C. Okeahialam, The effectiveness of antibacterial curtains in comparison with standard privacy curtains against transmission of microorganisms in a hospital setting, Le infezioni in medicina: rivista periodica di eziologia, epidemiologia, diagnostica, clinica e terapia delle patologie infettive, 27 (2019) 149-154.
[27] K. Shek, R. Patidar, Z. Kohja, S. Liu, J.P. Gawaziuk, M. Gawthrop, A. Kumar, S. Logsetty, Rate of contamination of hospital privacy curtains in a burns/plastic ward: A longitudinal study, Am. J. Infect. Control, 46 (2018) 1019-1021.
[28] B. Standard, Acoustics-determination of sound absorption coefficient and impedance in impedance tubes-Part 2: Transfer Function Method, in: BS EN ISO 10534-2:2001, 2001, pp. 10534-10532.

[29] K.P. Weilnau, E.R. Green, Z.E. Lampert, J.T. Kunio, Measurement of Small Fabric Samples using the Transmission Loss Tube Apparatus, in: Noise-Con 2016, Providence, Rhode Island, 2016.
[30] ASTM E90, Standard test method for laboratory measurement of airborne sound transmission loss of building partitions and elements, in: ASTM International: West Conshohocken, PA, USA, 2009.
[31] ASTM E2235-04, Standard Test Method for Determination of Decay Rates for Use in Sound Insulation Test Methods, in: ASTM International: West Conshohocken, PA, USA, 2012.
[32] American National Standard, ANSI S1.11: Specification for Octave-band and Fractional-octave-band Analog and Digital Filters, in, Standards Secretariat, Acoustical Society of America, 2004.
[33] B. Berglund, T. Lindvall, D.H. Schwela, New WHO guidelines for community noise, noise & vibration worldwide, 31 (2000) 24-29.
[34] G. Loupa, A. Katikaridis, D. Karali, S. Rapsomanikis, Mapping the noise in a Greek general hospital, Science of the Total Environment, 646 (2019) 923-929.
[35] M. Livera, B. Priya, A. Ramesh, P.S. Rao, V. Srilakshmi, M. Nagapoornima, A. Ramakrishnan, M. Dominic, Spectral analysis of noise in the neonatal intensive care unit, The Indian Journal of Pediatrics, 75 (2008) 217.
[36] X. Wang, F. You, F.S. Zhang, J. Li, S. Guo, Experimental and theoretic studies on sound transmission loss of laminated mica-filled poly (vinyl chloride) composites, J. Appl. Polym. Sci., 122 (2011) 1427-1433.
[37] J. Llorens, Fabric structures in architecture, 1st ed., Woodhead Publishing, Elsevier, 2015.
[38] M. Long, Architectural acoustics, 1st ed., Academic Press, Elsevier, 2005.
[39] H.C.P. Cordero, R.Q. ABARCA, K.D. Vargas, I.P. Troncoso, Polymeric materials with antifouling, biocidal, antiviral and antimicrobial properties; elaboration method and its uses, in, US Patents, 2016.